\documentclass[pra,twocolumn,showpacs,superscriptaddress,amsmath,amssymb]{revtex4}

\usepackage{graphicx}

\usepackage{color}

\begin{document}

\title{Bulk-edge correspondence, spectral flow and  Atiyah-Patodi-Singer theorem \\ 
for the $\mathcal{Z}_2$-invariant in topological insulators}
\author{Yue Yu}
\affiliation{ Center for Field Theory and Particle Physics, Department of Physics, and Key Laboratory of Surface Physics, Fudan University, Shanghai 200433, China}
\affiliation {Collaborative Innovation Center of Advanced Microstructures, Nanjing 210093, China}
\author{Yong-Shi Wu}
\affiliation{Department of Physics and Astronomy, University of Utah, Salt Lake
City, UT 84112, USA}
\affiliation{ Center for Field Theory and Particle Physics, Department of Physics, and Key Laboratory of Surface Physics, Fudan University, Shanghai 200433, China}
\affiliation {Collaborative Innovation Center of Advanced Microstructures, Nanjing 210093, China}
\author{Xincheng Xie}
\affiliation{International Center for Quantum Materials, School of Physics, Peking University, Beijing 100871, China}
\affiliation{Collaborative Innovation Center of Quantum Matter, Beijing 100871, China}


\date{\today}
\begin{abstract}
We study the bulk-edge correspondence in topological insulators by taking Fu-Kane spin pumping model as an example.  We show that the Kane-Mele invariant  in this model is $\mathcal{Z}_2$ invariant modulo the spectral flow of a single-parameter
family of 1+1-dimensional Dirac operators with a global boundary 
condition induced by the Kramers degeneracy of the system. This 
spectral flow is defined as an integer which counts the difference 
between the number of eigenvalues of the Dirac operator family 
that flow from negative to non-negative and the number of eigenvalues 
that flow from non-negative to negative. Since the bulk states of 
the insulator are completely gapped and the ground state is assumed 
being no more degenerate except the Kramers, they do not contribute 
to the spectral flow and only edge states contribute to. The parity 
of the number of the Kramers pairs of gapless edge states is exactly
the same as that of the spectral flow. This reveals the origin of the edge-bulk correspondence, i.e., why
the edge states can be used to characterize the topological insulators.
Furthermore, the spectral flow is related to the reduced 
$\eta$-invariant and thus counts both the discrete ground state
degeneracy and the continuous gapless excitations, which
distinguishes the topological insulator from the conventional band
insulator even if the edge states open a gap due to a strong interaction between  edge
modes. We emphasize that these results are also valid
even for a weak disordered and/or weak interacting system. The higher spectral flow to categorize 
the higher-dimensional topological insulators are expected.

\end{abstract}

 \maketitle

\section{Introductions}

Studies on the topological classification of the new condensed
matter systems have greatly enriched the zoo of quantum states of
matter \cite{ram,ktheory,ram1}. Pioneer development in this field was
the discovery of the topological nature of the quantum Hall effect
which is described by the TKNN topological invariant for a pure
system \cite{TKNN} and its generalization to disordered and
interacting systems \cite{NT,NWT}.

Recently found topological insulators with time reversal symmetry \cite{rev1,rev2} make
 the family of the topological quantum matter become very
affluent \cite{KM1,KM2,BZ}. These systems may be robust to the
weak disorders and weak interactions \cite{Wu,Xu,qwz,essin,shen,xie,groth}.
Experimentally, these time reversal invariant insulators have been
or are possibly realized in quantum wells, alloys and crystals
\cite{exp1,exp2,exp3,exp4,exp5,exp6,exp7}.

These time reversal invariant systems own a common topological
property, i.e., a $\mathcal{Z}_2$-valued topological invariant
\cite{KM1,KM2}, instead of the integer-valued one like the TKNN
invariant. We name this $\mathcal{Z}_2$-valued topological
invariant as Kane-Mele's invariant. Correspondingly, these
 systems with non-trivial  Kane-Mele's invariant  are of different edge states from
the conventional band insulator: The odd number of the pairs of
the gapless edge excitations \cite{KM1, KM2}.

On the other hand, the topological insulator can also be
characterized by another kind of topological invariant:
Descendence to $3+1$- and $2+1$-dimensions from the Chern-Simons
in a $4+1$-dimensional space-time which includes the
three-dimensional Brillouin zone \cite{QHZ}.

These two different topological invariants are proved being
exactly equivalent \cite{wqz}. In fact, this equivalence gives an
index theorem between the topological index which is discrete
(i.e., Kane-Mele's here) and an analytic index (i.e., the
Chern-Simons and its descendence), similar to the Euler number,
which is a topological index defined by the triangulation, is
exactly equal to the analytic index given by Gauss-Bonnet theorem.
A more relevant index theorem is the index theorem of Dirac
operator because the gapless edge excitations of these topological
insulators are exactly Dirac-type.

Although the topological invariant defined over the Brillouin zone
may nicely reflect the topological property of the quantum space,
it is not clear how this topological invariant relates to the
topological index of the Dirac operator acting on the quantum
space in the real space representation. For the topological
$p$-wave paired superfluid, such a relation has been provided by
one of the authors \cite{Yue}. In the topological insulators, for
both with and without the time reversal symmetry, the definition
of the topological invariants stemming from the real space lacks.

The bulk-edge correspondence, more generally, the bulk-boundary correspondence, is an essential phenomenon in topological states of matter. 
This has been known since the integer quantum Hall effects were discovered \cite{halp}.  Many recent studies focus on this problem \cite{bec1,
bec2,bec3,bec4,bec5,bec6,bec7,bec8,bec9}. There are several expressions of this correspondence according to the problems in various scenarios. Topologically,  this  is the correspondence between the  number of the {\it real space} gapless edge (boundary) states and the {\it momentum space} topological invariant. It is a fact but not clear yet why such a correspondence hold.  In our opinion,    
to well understand this correspondence from a topological point of view, Atiyah-Patodi-Singer index theorem  \cite{APS,APS1}  for Dirac operator  must play a key role. 

To study such a correspondence for arbitrary dimensional space-time
requires some more mathematical tools that the condensed matter
physicists are not familiar. We will gradually introduce to
readers these tools in the present work and subsequent works. As
the first step, we study the systems with a $1+1$-dimensional
Dirac operator. An example of such a system was provided Fu and
Kane \cite{FK}. Its lattice version is a one-dimensional tight
binding model of electrons with a staggered magnetic field, a
staggered bond modulation as well as a time reversal invariant
spin-orbit coupling. For this model, Fu and Kane analyzed the
$\mathcal{Z}_2$ adiabatic spin pump between the topological
insulating material and a reservoir. The $\mathcal{Z}_2$-valued
topological invariant in this $\mathcal{Z}_2$ pump is similar to
the $\mathcal{Z}_2$ topological invariant (Kane-Mele's invariant)
in a two-dimensional topological insulator. In the continuous
limit, there is a well-defined Dirac operator in this model. We
can show that its spectral problem can be solved with a so-called
Atiyah-Patodi-Singer global boundary condition \cite{APS,APS1}, which
is induced by the Kramers degeneracy.

With this example in mind, we study a model-independent
Atiyah-Patodi-Singer boundary problem for a 1+1-dimensional
topological insulator. We show that Kane-Mele's invariant is
$\mathcal{Z}_2$-modular to the spectral flow of a single-parameter
family of the Dirac operators. Here, the spectral flow is defined
as {\it an integer which counts the difference between the number
of eigenvalues of the Dirac operator family that flow from
negative to non-negative and the number of eigenvalues that flow
from non-negative to negative}. Therefore, Kane-Mele's invariant
is computed by counting the discrete ground state degeneracy and
the number of the continuous gapless excitations.

In a pure topological insulator where the Dirac operators can be
parameterized by the Brillouin zone, the ground state is assumed
to be not degenerate except Karmers degeneracy and thus the parity
of the spectral flow is the same as that of the number of the
pairs of the gapless edge states. This explains why the edge
states can distinguish the topological insulator from the
conventional band insulator.

Since the spectrum of the Dirac operators is determined by the
topology of the quantum state space based on the real space-time,
we can also alter the parametrization other than the Brillouin
zone when it is not a convenient one. For a system with disorder
or interaction, a well-known parametrization is the twisted
boundary condition \cite{NT,NWT}, whose usage to the
1+1-dimensions \cite{FK} and other topological insulator systems
\cite{essin} have been studied. Thus, our results can also be used
to the 1+1-dimensional topological insulator with disorder and
weak interaction.

For a strong interaction between the edge states, a gap opens at
the edge  but the ground state becomes degenerate \cite{Wu,Xu,FK}
while the topological nature described by Kane-Mele's invariant is
still kept. There is a theorem that a non-trivial topological
invariant may lead to either ground state degeneracy or gapless
excitations \cite{qwz}. From the spectral flow point of view, it
is easy to be understood because the spectral flow counts both the
discrete ground state degeneracy and the number of the continuous
gapless excitations.

The strong interaction between the bulk Dirac particles may modify the 
$Z_2$ topological insulators to $Z_8$ ones \cite{citi}. Since the Atiyah-Patodi-Singer boundary problem concerns the linear Dirac equation, 
our result can not be directly applied to the strong interacting topological insulators.  More mathematic tools concerning the 
non-linear Dirac equations are required and this is out of our goal in this paper.

The studies for the 1+1-dimensional topological insulators can be
generalized to higher-dimensional space-time but this involves
some advanced mathematical tools, e.g., to define a higher
spectral flow \cite{DaiZhang}. We shall sketch the possible
generalization and leave the details for the subsequent studies.

This paper is organized as follows: In Sec. II, we introduce the
time reversal invariant model with $\mathcal{Z}_2$ spin pump given
by Fu and Kane \cite{FK} and its continuous limit. The
Atiyah-Patodi-Singer boundary condition problem corresponding to
the Kramers degeneracy is set up. In Sec. III, we define the
$\eta$-invariant and spectral flow \cite{KL}. In Sec. IV, we prove
a theorem that relates the $\eta$-invariant to the boundary
projection that is defined by the Atiyah-Patodi-Singer boundary
condition. Mathematically, it is the Scott-Wojciechowski theorem
\cite{SW}. The section V devotes our main result, Kane-Mele's
invariant, a winding number related to the boundary projection, is
$\mathcal{Z}_2$-modular to the spectral flow of the family of the
Dirac operators. In Sec. VI, we generalize the main result to
disordered and interacting systems. In Sec. VII, we argue that our
result may be generalized to the higher-dimensional space-time.
The final section is discussions and conclusions.

\section{1+1-dimensional topological insulator and Atiyah-Patodi-Signer boundary problem}

Although the Atiyah-Patodi-Signer boundary problem we shall solve
is not dependent on a concrete model, having such a system is
helpful to obtain an intuitive understanding. In this section, we
first review the 1+1-dimensional topological insulator model with
$\mathcal{Z}_2$ spin pump proposed by Fu and Kane and Kane-Mele's
invariant in this model \cite{FK}. In the continuous limit, we
obtain a single-parameter family of the Dirac operators in
1+1-dimensional space-time. The first three subsections basically
are the summary of those results we will use in this work. With
the Kramers degeneracy, we then abstract a model-independent
Atiyah-Patodi-Signer boundary problem which is the starting point
to show the relation between Kane-Mele's invariant and the
spectral flow of the Dirac operator family.

\subsection{Tight Binding Model}

Consider a one-dimensional lattice in which electrons are allowed
only hopping between the nearest neighbor sites and suffer a
staggered magnetic field and a staggered bond modulation as well
as a time reversal invariant spin-orbit coupling. Namely, the
Hamiltonian of the system is given by
\begin{eqnarray}
 H=H_0 + V_h + V_b+ V_{so}, \label{lv}
 \end{eqnarray}
where
\begin{eqnarray}
H_0 &=& -t_0\sum_{\langle ij\rangle,\alpha}(c^\dag_{i\alpha}c_{j\alpha}+h.c.),\nonumber\\
V_h&=& h\sum_{i,\alpha\beta}(-1)^i\sigma_{\alpha\beta}^z
 c^\dag_{i\alpha}c_{i\beta},\nonumber\\
V_b&=& b\sum_{\langle ij\rangle,\alpha}
(-1)^i(c^\dag_{i\alpha}c_{j\alpha}+h.c),\nonumber\\
V_{so}&=&\sum_{\langle ij\rangle,\alpha\beta} i{\bf e}\cdot{\vec
\sigma}_{\alpha\beta}(c^\dag_{i\alpha}c_{j\beta}-c^\dag_{j\alpha}c_{i\beta}),\nonumber
\end{eqnarray}
The last term is the spin-orbital coupling which is characterized
by the vector ${\bf e}$ and violates the conservation of $S_z$. Fu
and Kane consider an adiabatic cycle as time $t$ varying from 0 to
$T$ \cite{FK}
\begin{eqnarray}
b(t)=b^0\cos(2\pi t/T),~h(t)=h^0\sin(2\pi t/T).
\end{eqnarray}
 The Hamiltonian becomes time reversal invariant only
at $t=0$ and $t=T/2$.

The spectrum has been calculated and turns out the existence of a
single pair of gapless edge excitations. This characterizes this
non-trivial topological insulator.

\subsection{Kane-Mele's Invariant}

We assume the one-dimensional lattice has a lattice constant $a=1$
and the length $L$ with a periodic boundary condition. The $2N$
bands are occupied.  A unitary transformation
$H(k)=e^{ikx}He^{-ikx}$ parameterizes the Hamiltonian in the
Brillouin zone $-\pi\leq k\leq \pi$. Under a time reversal
transformation,
\begin{eqnarray}
\Theta H(-k,-t)\Theta^{-1}=H(k,t),
\end{eqnarray}
with the time reversal operator
\begin{eqnarray}
\Theta=e^{i\pi S_y} C
\end{eqnarray}
for $S_y$ is the $y$-component of the spin operator and $C$ being
the complex conjugate. The spectrum is Kramers degenerate at
$(k,t)=\{\Gamma_i\}=\{(0,0),(0,T/2),(\pi, 0),(\pi,T/2)\}$ since
$H(k,t)$ is time reversal invariant at these points.

According to the Bloch theorem, the $n$-band Bloch wave function
reads
\begin{eqnarray}
|\psi_{k,n}\rangle=\frac{1}{\sqrt L}e^{ikx}|u_{k,n}\rangle
\end{eqnarray}
where $|u_{k,n}\rangle$ is cell-periodic.

Due to the Kramers degeneracy, the time reversal operator defines
a map $w$ from $|u_{k,n}\rangle$ to its Kramers partner
$|u_{m,-k}\rangle$, namely,
\begin{eqnarray}
w_{mn}=\langle u_{-k,m}|\Theta|u_{k,n}\rangle\label{map}
\end{eqnarray}
In fact, also due to the Kramers degeneracy, the $2N$ bands can be
labelled by $(\alpha,a)$ for $\alpha=1,\cdots,N$ and $a=0,1$. For
a given $\alpha$, the eigenvalues of the Hamiltonian
$E_\alpha^0(-k)=E_\alpha^1(k)$ due to time reversal symmetry.
Therefore, up to a $U(1)$ phase, $|u_{-k,\alpha}^0\rangle$ is the
Kramers partner of $|u_{k,\alpha}^1\rangle$ \cite{FK},
\begin{eqnarray}
&&|u_{-k,\alpha}^0\rangle=e^{i\chi_{k,\alpha}}\Theta|u_{k,\alpha}^1\rangle,
\nonumber\\
&&|u_{-k,\alpha}^1\rangle=-e^{i\chi_{-k,\alpha}}\Theta|u_{k,\alpha}^0\rangle,
\label{project}\\
&&\langle u^a_{-k,\alpha}|=(-1)^a\langle
u^{a+1}_{k,\alpha}|\Theta^\dag
e^{-i\chi_{(-1)^{a}k,\alpha}}.\nonumber\\
\nonumber
\end{eqnarray}
The effective parameter space $\tau_1$  in $(k,t)$ is reduced to
the area that is surrounded by the lines linking four time
reversal invariant points $\Gamma_i$. Kane-Mele's
$\mathcal{Z}_2$-valued invariant which categorizes this
topological insulator is given by
\begin{eqnarray}
{\rm
wind}(w)_{\partial\tau_1}=\sum_i[\log\sqrt{\det(w(\Gamma_i))}-\log{\rm
Pf}(w(\Gamma_i))],\label{wind1}
\end{eqnarray}
which is the winding number of the mapping $w$ around the boundary
$\partial \tau_1$ of the effective parameter space $\tau_1$.

\subsection{Continuous Limit}

To relate to the Atiyah-Patodi-Singer boundary problem, we go to
the continuous limit of the lattice model. The continuous version
of (\ref{lv}) is described by the Hamiltonian density \cite{FK}
\begin{eqnarray}
H&=&\psi^*_{a\alpha}(iv_F\tau^z_{ab}\delta_{\alpha\beta}\partial_x
+h\tau^x_{ab}\sigma^z_{\alpha\beta}\nonumber\\
&+&b\tau^y_{ab}\delta_{\alpha\beta}+\tau^z_{ab}{\bf e}\cdot{\vec
\sigma}_{\alpha\beta})\psi_{b\beta}
\end{eqnarray}
where $\psi_{a\alpha}$ is four-component fermion field with
$a=L,R$ denoting the left- and right-movings and $\alpha$ being
the spin. $\tau$ and $\sigma$ are Pauli matrices corresponding to
$a$ and $\alpha$, respectively. From this Hamiltonian, we read out
an effective Dirac operator
\begin{eqnarray}
D(t)=iv_F\tau^z \sigma^0\partial_x +h(t)\tau^x\sigma^z
+b(t)\tau^y\sigma_0+\tau^z{\bf e}\cdot{\vec \sigma},
\end{eqnarray}
where $\sigma^0$ is the identity matrix acting on the spin.

\subsection{The Atiyah-Patodi-Singer Boundary Problem}

We have presented an example of 1+1-dimensional topological
insulator with time reversal invariant. There is a Dirac operator
which relates to the gapless edge Dirac cone. Parameterizing the
Dirac operator with the Brillouin zone,
$D(k,t)=e^{-ikx}D(t)e^{ikx}$, we have a family of the Dirac
operators.

For a given Dirac operator in a space-time with a boundary, a
local boundary condition is often not allowed \cite{APS,APS1}. Instead,
one has to introduce a global boundary condition on the boundary.

To describe a global boundary condition, we first make some
definitions.

(i) Let Spec$D=\{\lambda\}$ denote all the eigenvalues that obey
$D\psi_\lambda=\lambda\psi_\lambda$. We call Spec$D$ the spectrum
of $D$.

(ii) Let Spec$_PD=\{\lambda_P\}$ be a subset of Spec$D$.

(iii) Denote
$|\phi_P\rangle=\sum_{\lambda_P}a_{\lambda_P}|\psi_{\lambda_P}\rangle$.
Define a projection $P$ so that $P|\phi_P\rangle=0$ for all
$|\phi_P\rangle$.

For a space-time $M$ and its boundary $\partial M$, a global
boundary condition is often given by, for a projection $P$,
$P|\phi_P\rangle|_{\partial M}=0$.

A general  Atiyah-Patodi-Singer boundary problem is to solve the
problem $D|\psi_\lambda\rangle=\lambda|\psi_\lambda\rangle$ with a
global boundary condition $P|\phi_P\rangle_{\partial M}=0$ for a
subset of $\{\lambda_P\}$. For simplicity, we denote the operator
$D_P$ the Dirac operator obeying the boundary condition.

For a family of the Dirac operators $D(k)$, we can have a family
of the boundary condition problems with $D_{P(k)}$.

The general description of the Atiyah-Patodi-Singer boundary
problem is somewhat abstract.  We present an example according to
the model we are studying in this section. The space-time
$\{(x,t)\}$ in the periodic boundary condition is a
two-dimensional torus. Due to the Kramers degeneracy, we study a
cylinder $M=S^1\times [0,T/2]$. The boundary of this cylinder is
$\partial M =(S^1\times\{0\})\cup (S^1\times\{T/2\})$.  For the
family of the $D(k)$ with the wave functions
$|u^{a}_{k,\alpha}\rangle$, we define a map $Q(k)$:
\begin{eqnarray}
Q(k)|u^{a}_{-k,\alpha}\rangle=|u^{a}_{k,\alpha}\rangle, \label{Q}
\end{eqnarray}
The Kramers degeneracy related to the time reversal symmetry
defines a projection , i.e., we can rewrite  (\ref{project}) as
\begin{eqnarray}
&&P(k)\left[\begin{array}{c}
                   |u^0_{-k,\alpha}\rangle\\
                   |u^1_{-k,\alpha}\rangle\\
                   \end{array}\right]\label{proj}\\
                   &&=\frac{1}2\left[\begin{array}{c}
                   |u^0_{-k,\alpha}\rangle-e^{i\chi_{k,\alpha}}\Theta Q(k)|u^{1}
                   _{-k,\alpha}\rangle\\
                   |u^1_{-k,\alpha}\rangle+e^{i\chi_{-k,\alpha}}\Theta Q(k)|u^{0}
                   _{-k,\alpha}\rangle
                   \end{array}\right]=0.\nonumber
\end{eqnarray}

The Atiyah-Patodi-Singer boundary problem is now defined by
\begin{eqnarray}
D(k)\psi(x,t)|_{\partial M}=\lambda(k)\psi(x,t)|_{\partial M}
\end{eqnarray}
with the global boundary condition 
\begin{eqnarray}
P(k)\psi|_{\partial M}=0. \label{APSBC}
\end{eqnarray}
We said the boundary condition (\ref{APSBC}) is global because the project operator $P(k)$ is dependent on the wave vector $k$ but not locally on the coordinate $x$. 

This is the problem we are going to solve in the rest of this
paper.  We would like to emphasize that the "boundary condition" (\ref{APSBC}) in this Atiyah-Patodi-Singer boundary
problem is not the real physical boundary condition of the system in which the wave function is periodic both in the spatial and time directions.  In the materials, the physical boundary condition may be more different from (\ref{APSBC}). For example, instead of on the circle $S^1$, the one-dimensional lattice in reality is a chain with open boundaries.     
However, the physical boundary changes do not affect our results as we will argued later. (See Sec. V B.) 

\section{$\eta$-invariant and spectral flow}

As we have introduced, to relate Kane-Mele's invariant to the edge
states, one needs to use the concept of the spectral flow. In this
section, we define the spectral flow and its analytical expression
in the $\eta$-invariant. With the notations that are easier to be
understood by physicists, this section basically is a review to
the definitions and  results in Sec. 3 in Ref. \cite{KL}.

\subsection{$\eta$- and $\zeta$- invariants}

For the Dirac operator $D_P$, the $\eta$- and $\zeta$-functions
are defined, for Re$(s)\gg 0$, by
\begin{eqnarray}
\eta(D_P;s)&=&{\rm tr}(D_P|D_P|^{-s-1})=\sum_{\lambda\in {\rm
Spec}'
D_P} {\rm sign}(\lambda)|\lambda|^{-s},\nonumber\\
\zeta(D_P;s)&=&{\rm tr}(D_P^{-s})=\sum_{\lambda\in {\rm Spec}'
D_P} \lambda^{-s},\label{eta}
\end{eqnarray}
where the prime on Spec$D$ means the zero eigenvalues of $\lambda$
are excluded.

For a general $P$, one has
\begin{eqnarray}
\zeta(D_P;s)&=&\frac{1}2(\zeta(D_P^2;s/2)+\eta(D_P;s)),\label{etazeta}
\\
&+&(-1)^{-s}\frac{1}2(\zeta(D_P^2;s/2)-\eta(D_P;s)),\nonumber
\end{eqnarray}
which can be seen by writing the right hand side as
\begin{eqnarray}
&&\frac{1}2[\sum_{\lambda_P>0}((\lambda_P^2)^{-s/2}+\lambda_P^{-s})\nonumber\\
&&+\sum_{\lambda_P<0}((|\lambda_P|^2)^{-s/2}-(-1)^{-s}|\lambda_P|^{-s})]\nonumber\\
&&+(-1)^{-s}\frac{1}2[\sum_{\lambda_P>0}((\lambda_P^2)^{-s/2}-\lambda_P^{-s})\nonumber\\
&&+\sum_{\lambda_P<0}((|\lambda_P|^2)^{-s/2}+(-1)^{-s}|\lambda_P|^{-s})]\nonumber\\
&&=\sum_{\lambda_P>0}\lambda_P^{-s}+\sum_{\lambda_P<0}(-1)^s|\lambda_P|^{-s}.\nonumber
\end{eqnarray}
The right hand side of the last equality is exactly the definition
of $\zeta(D_P;s)$.

For the projection $P=P(k)$ \cite{note}, the $\eta$- and
$\zeta$-functions are regular at $s=0$. Moreover,
$\zeta(D_{P(k)};0)$ is not dependent on $k$.

The $\eta$-invariant of $D_P$ is defined by
\begin{eqnarray}
\eta(D_P)=\eta(D_P;0),
\end{eqnarray}
which counts the asymmetry between non-zero positive eigen states
and non-zero negative eigen states. To include the zero modes, we
define so-called {\it reduced $\eta$-invariant}:
\begin{eqnarray}
\tilde\eta(D_P)=\frac{1}2(\eta(D_P)+\eta_0(D_P))
\end{eqnarray}
where $\eta_0(D_P)$ is the number of the independent  zero modes.

For a general projection $P$ , a family of $\eta(D_{P(k)})$ does
not vary smoothly in the whole interval $k\in [0,\pi]$. However,
$\tilde\eta(D_{P(k)})$ is smooth \cite{grubb}. Furthermore, a
real-valued function or a $k$-dependent Berry phase, is smooth,
\begin{eqnarray}
B(k)=\int _0^kdk' \frac{d}{dk'}\eta(D_{P(k')}).
\end{eqnarray}
 The above mathematical results are consistent with the
existence of $\mathcal{Z}_2$ obstacle found by Fu and Kane
\cite{FK,note2}.

\subsection{Spectral Flow}

We now study the relation between spectral flow and
$\eta$-invariant. For the family of $D(k)$ with $k\in [0,\pi]$
which obeys (\ref{proj}), the family of eigenvalues $\lambda_P(k)$
may or may not cross the Fermi level (which is set to be zero).
Define a direction of the curve $\lambda_P(k)$ which starts from
$\lambda_P(0)$ and ends at $\lambda_P(\pi)$. A curve starting from
$\lambda_P(0)<0$ and ending at $\lambda_P(\pi)\geq 0$ is endowed a
number $+1$ while a curve starting from $\lambda_P(0)\geq 0$ and
ending at $\lambda_P(\pi)<0$ is endowed $-1$. Other curves are
endowed 0. {\it Summation of all these number defines the spectral
flow of this Dirac operator family.}

For a given $k\in[0,\pi]$, assume $\pm
\epsilon~\bar\in~$Spec$D_{P(k)}$ with $\epsilon>0$. Then for a
small interval $k\in [k_0,k_1]$, $\eta(D_{P(k)};s)$ can be divided
into two parts
\begin{eqnarray}
\eta(D_{P(k)};s)&=&\sum_{0<|\lambda_p|<\epsilon}{\rm
sign}(\lambda_P)|\lambda_P|^{-s}\nonumber\\
&+&\sum_{|\lambda_P|>\epsilon}{\rm
sign}(\lambda_P)|\lambda_P|^{-s}\\
&=&\eta_{<\epsilon}(D_{P(k)};s)+\eta_{>\epsilon}(D_{P(k)};s),\nonumber
\end{eqnarray}
where $\eta_{<\varepsilon}(D_{P(k)};0)$ is a finite integer. One
then has
\begin{eqnarray}
&&\eta_{<\varepsilon}(D_{P(k_1)};0)
-\eta_{<\varepsilon}(D_{P(k_0)};0)=2{\rm SF}(D_P(t))_{k\in[k_0,k_1]}\nonumber\\
&&+{\rm zero~modes~of~}D_{P(k_0)}-{\rm
zero~modes~of~}D_{P(k_1)}\label{diff}
\end{eqnarray}
To understand this fact, we examine the following simple example.
Assuming two sets of eigenvalues
\begin{eqnarray}
\{\lambda_1(k_0)<0,\lambda_2(k_0)<0,
\lambda_3(k_0)=0,\lambda_4(k_0)=0, \lambda_5(k_0)>0\}\nonumber\\
\{\lambda_1(k_1)<0,\lambda_2(k_1)=0,\lambda_3(k_1)=0,\lambda_4(k_1)=0,
\lambda_5(k_1)>0 \},\nonumber
\end{eqnarray}
the direct counting  gives \[\eta_{<\varepsilon}(D_{P(k_1)};0)
-\eta_{<\varepsilon}(D_{P(k_0)};0)=(-1+1)-(-2+1))=1,\] while
dimension difference of the zero modes is $-1$ and 2SF=$2\times
1=2$, which also gives $1$.

Thus, we have
\begin{eqnarray}
\eta_{>\epsilon}(D_{P(k)});s)&=&\eta(D_{P(k)};s)-\eta_{<\epsilon}(D_{P(k)};s)\nonumber\\
&=& \eta(D_{P(k)}){\rm mod}~\mathcal{Z},\nonumber
\end{eqnarray}
 which varies smoothly in
$k\in [k_0,k_1]$. The Berry phase between $[k_0,k_1]$ is given by
\begin{eqnarray}
&&\int_{k_0}^{k_1}dk\frac{d}{dk}\eta(D_{P(k)})
=\int_{k_0}^{k_1}dt\frac{d}{dk}\eta_{>\epsilon}(D_{P(k)})\nonumber\\
&&=\eta_{>\epsilon}(D_{P(k_1)})-\eta_{>\epsilon}(D_{P(k_0)})
=\eta(D_{P(k_1)})-\eta(D_{P(k_0)})\nonumber\\
&&-\eta_{<\epsilon}(D_{P(k_1)})+\eta_{<\epsilon}(D_{P(k_0)})
\end{eqnarray}
According to (\ref{diff}), one has
\begin{eqnarray}
\int_{k_0}^{k_1}dk\frac{d}{dk}\eta(D_{P(k)})
&=&\eta(D_{P(k_1)})-\eta(D_{P(k_0)})\nonumber\\
&-&2{\rm SF}(D_P(t))_{k\in[k_0,k_1]}\\
&-&{\rm zero~modes~of~}D_{P(k_0)}\nonumber\\
&+&{\rm zero~modes~of~}D_{P(k_1)}\nonumber
\end{eqnarray}

 Dividing the interval $[0,\pi]$ into small intervals
$[k_i,k_{i+1}]$ and  summation over all these small intervals, we
get a relation between the $\eta$-invariant and the spectral flow
\begin{eqnarray}
\tilde\eta(D_{P(\pi)})-\tilde\eta(D_{P(0)})&=&{\rm
SF}(D_{P(k)})_{k\in [0,\pi]}\nonumber\\
&+&\frac{1}2\int _0^\pi dk\frac{d}{dk}\eta(D_{P(k)}).\label{etasf}
\end{eqnarray}

\section{The Scott-Wojciechowski Theorem: an expression of $\eta$-invariant}

If we have the whole spectrum, the $\eta$-invariant and the
spectral flow can be directly read out by examining the sign of
each eigenvalue as well as number of the zero modes. However, our
task is to relate them with a topological invariant winding
number. That is, we need to relate them to the map $w$ given by
the projection $P(k)$. The Scott-Wojciechowski theorem states a
relation between $\zeta$-function and the determinant of the map
$w$ associated with the projection $P$\cite{SW}. Using Eq.
(\ref{etazeta}), we then set up a relation between
$\eta$-invariant and the determinant of $w$.

\subsection{{\it w} Map}

We first rewrite the projection $P$. The map $Q$ (Eq. (\ref{Q}))
corresponds to a projection $P_Q$,
\begin{eqnarray}
&&0=P_Q\left[\begin{array}{c}
                   u_{k,\alpha}^0\\
                   u_{k,\alpha}^1\\
                   u_{-k,\alpha}^0\\
                   u_{-k,\alpha}^1
                   \end{array}\right]=\frac{1}2\left[\begin{array}{c}
                   u_{k,\alpha}^0-Q u_{-k,\alpha}^0\\
                   u_{k,\alpha}^1-Qu_{-k,\alpha}^1\\
                   u_{-k,\alpha}^0-Qu_{k,\alpha}^0\\
                   u_{-k,\alpha}^1-Qu_{k,\alpha}^1
                   \end{array}\right]\nonumber\\&&=\frac{1}2\left[\begin{array}{cccc}
                   1&0&-Q&0\\
                   0&1&0&-Q\\
                   -Q&0&1&0\\
                   0&-Q&0&1
                   \end{array}\right]\left[\begin{array}{c}
                   u_{k,\alpha}^0\\
                   u_{k,\alpha}^1\\
                   u_{-k,\alpha}^0\\
                   u_{-k,\alpha}^1
                   \end{array}\right]\nonumber\\
                   &&=\frac{1}2\left[\begin{array}{cc}
                   I&\Phi(P_Q]\\
                   \Phi(P_Q)&I\end{array}\right]\left[\begin{array}{c}
                   u_{k,\alpha}\\
                   u_{-k,\alpha}\end{array}\right]
                   \end{eqnarray}
and the map $P(k)$ (Eq.(\ref{proj})) can be rewritten as
\begin{eqnarray}
0&&=P_\Theta\left[\begin{array}{c}
                   u_{k,\alpha}^0\\
                   u_{k,\alpha}^1\\
                   u_{-k,\alpha}^0\\
                   u_{-k,\alpha}^1
                   \end{array}\right]
                   =\frac{1}2\left[\begin{array}{c}
                   u_{k,\alpha}^0-e^{i\chi_{-k,\alpha}}\Theta u_{-k,\alpha}^1\\
                   u_{k,\alpha}^1+e^{i\chi_{k,\alpha}}\Theta u_{k,\alpha}^0\\
                   u_{-k,\alpha}^0-e^{i\chi_{k,\alpha}}\Theta u_{k,\alpha}^1\\
                   u_{-k,\alpha}^1+e^{i\chi_{-k,\alpha}}\Theta u_{k,\alpha}^0
                   \end{array}\right]\nonumber\\
                   =&&
                   \frac{1}2\left[\begin{array}{cccc}
                   1&0&0&-e^{i\chi_{-k,\alpha}}\Theta\\
                   0&1&e^{i\chi_{k,\alpha}}\Theta&0\\
                   0&-e^{i\chi_{k,\alpha}}\Theta&1&0\\
                   e^{i\chi_{-k,\alpha}}\Theta&0&0&1
                   \end{array}\right]\nonumber\\&&\times\left[\begin{array}{c}
                   u_{k,\alpha}^0\\
                   u_{k,\alpha}^1\\
                   u_{-k,\alpha}^0\\
                   u_{-k,\alpha}^1
                   \end{array}\right]
                  = \frac{1}2\left[\begin{array}{cc}
                   I&\Phi(P_\Theta,-k)\\
                   \Phi(P_\Theta,k)&I\end{array}\right]\left[\begin{array}{c}
                   u_{k,\alpha}\\
                   u_{-k,\alpha}\end{array}\right]\nonumber
                   \end{eqnarray}
At $\Gamma_i$, note that $\Theta^\dag=-\Theta$ and
$e^{ia}\Theta=-\Theta^\dag e^{-ia}$, $Q=1$ and $P_\Theta$ is
hermitian, i.e., $\Phi(P_\Theta,-k)=\Phi^\dag(P_\Theta,k)$.
Therefore, at $\Gamma_i$,
\begin{eqnarray}
 &&\Phi(P_Q)\Phi^\dag(P_\Theta)=\left[\begin{array}{cc}
      -Q&0\\
      0&-Q
      \end{array}\right]\left[\begin{array}{cc}
      0&-e^{i\chi_{k,\alpha}}\Theta\\
      e^{i\chi_{k,\alpha}}\Theta&0
      \end{array}\right]\nonumber\\
      &&=\left[\begin{array}{cc}
      0&e^{i\chi_{k,\alpha}}\Theta\\
      -e^{i\chi_{k,\alpha}}\Theta&0
      \end{array}\right]\nonumber
      \end{eqnarray}

On the other hand, rearranging
\[(u_{k,m})=(u^0_{k,1},u^1_{k,1},\cdots,u^0_{k,N},u^1_{k,N}),\]
 the map (\ref{map}) reads
\begin{eqnarray}
w=\left[\begin{array}{cccccc}
        0&e^{i\chi_{k,1}}&0&\cdots&0&0\\
        -e^{i\chi_{-k,1}}&0&\cdots&0&0&0\\
        ~&~&~&\cdots&~&~\\
        0&0&0&\cdots&0&e^{i\chi_{k,N}}\\
        0&0&\cdots&0&-e^{i\chi_{-k,N}}&0
        \end{array}\right]
        \end{eqnarray}
Thus, the map $w$ is equivalent to the map $\Phi$.  At $\Gamma_i$,
this is antisymmetric and
\begin{eqnarray}
{\rm Pf}[\Phi(P_Q)\Phi^\dag(P_\Theta)(\Gamma_i)]={\rm
Pf}[w(\Gamma_i)]=e^{i\sum_\alpha \chi_{k,\alpha}}
\end{eqnarray}

With these equations in mind, we go to the Scott-Wojciechowski
theorem.

\subsection{The Scott-Wojciechowski Theorem and Expression of $\eta$-invariant}

 For a smooth $P$, e.g., $P(k)$ we are considering, define a $\zeta$-determinant
 by
\begin{eqnarray}
{\rm det}_\zeta(D_P)=\left\{\begin{array}{l}
e^{-\zeta'(D_P,0)},~0~\bar\in~
{\rm spec}D_P\\
0.~~~~~~~~~~~~0\in {\rm spec}D_P
\end{array}\right.
\end{eqnarray}
 where $\zeta'(D^2_{P};0))=\frac{\partial}{\partial s}\zeta(D^2_{P};s))|_{s\to 0}$.
This is called a determinant because $\zeta$-function is a kind of
trace.

 According to (\ref{etazeta}), the $\zeta$-determinant is
rewritten by
\begin{eqnarray}
{\det}_\zeta(D_P)=e^{\frac{1}2(\zeta(D^2_P,0)-\eta(D_P,0))-
\frac{1}2\zeta'(D_P^2,0)},\label{27}
\end{eqnarray}

The Scott-Wojciechowski theorem \cite{SW} is stated as follows:
For a smooth projection $P$ which is of the form
\begin{eqnarray}
P=\frac{1}2\left[\begin{array}{cc}I&\Phi^\dag(P)\\
                          \Phi(P)&I\end{array}\right],\nonumber
                          \end{eqnarray}
the following relation holds
\begin{eqnarray}
{ \det} _\zeta(D_P)={\det} _\zeta
(D_{P_C})\det(\frac{I+\Phi(P_C)\Phi^\dag(P)}2)
\end{eqnarray}
 where the second "$\det$" on the right hand side is the conventional determinant of an operator
 acting on the Hilbert space. $P_C$ is so-called the Caldron
projector which is a reference projection \cite{KL}. We do not
give its definition and discuss its property because we will have
a better reference projection in the present case. Also we do not
explain the smooth projection. $P(k)$ we are considering are
smooth.

We would like to use this theorem to express the $\eta$-invariant
\cite{KL}. For the smooth projection $P$, $\zeta(D^2_P,0)$ is
independent of the concrete form of $P$. Thus, in terms of (\ref{27}) and assuming
$D_P$ has no zero modes \cite{KL}, one has
\begin{eqnarray}
&&e^{i\frac{\pi}2(\eta(D_{P_C}-\eta(D_P))}
e^{\frac{1}2(\zeta'(D^2_{P_C};0)-\zeta'(D^2_{P};0))}=\frac{
\det_\zeta(D_P)}{\det_\zeta(D_{P_C})}\nonumber\\
&&=\det(\frac{I+\Phi(P_C)\Phi^\dag(P)}2).
\end{eqnarray} 
Notice that the first factor in the left hand side of the first
equity is an pure phase and the second factor is real, i.e., the
product of two factors can be thought as $e^{i arg(\det)}| \det|$,
i.e.,
\begin{eqnarray}
\frac{\det(\frac{I+\Phi(P_C)\Phi^\dag(P)}2)}
{|\det(\frac{I+\Phi(P_C)\Phi^\dag(P)}2)|}=e^{i\frac{\pi}2(\eta(D_{P_C})-\eta(D_P))},
\label{det}
\end{eqnarray}

Since $\Phi(P)$ is a unitary operator, one can write
$\Phi(P_C)\Phi^\dag(P)=e^{iS}$ for $S^\dag=S$. Then
\begin{eqnarray}
&&\det(\frac{I+\Phi(P_C)\Phi^\dag(P)}2)^2=
\det(\frac{I+e^{iS}}2)^2\nonumber\\
&&=\det(e^{iS}\cos^2S/2)=\det(e^{iS})
\det(\cos^2S/2)\nonumber\\
&&=\det(\Phi(P_C)\Phi^\dag(P))\det(\cos^2S/2)
\end{eqnarray}
For a hermitian operator $S$, $\det(\cos^2S/2)$ is a positive real
number. The unitary of $e^{iS}$ means $|\det(e^{iS})|=1$ and $
\det(\Phi(P_C)\Phi^\dag(P))$ is only a phase. Therefore,
\begin{eqnarray}
\frac{\det(\frac{I+\Phi(P_C)\Phi^\dag(P)}2)^2}
{|\det(\frac{I+\Phi(P_C)\Phi^\dag(P)}2) |^2}=
\det(\Phi(P_C)\Phi^\dag(P)).\label{dets}
\end{eqnarray}
Comparing (\ref{det}) with (\ref{dets}), we have
\[e^{i\pi(\eta(D_{P_C})-\eta(D_P))}=\det(\Phi(P_C)\Phi^\dag(P))\]

In fact, this result works for an arbitrary smooth $P$ with $\eta$
replaced by $2\tilde\eta$ \cite{KL}. Therefore, for a pair of
smooth operators $(P,Q)$, one has

\begin{eqnarray}
e^{2\pi
i(\tilde\eta(D_{P})-\tilde\eta(D_Q))}=\det(\Phi(P)\Phi^\dag(Q)).
\end{eqnarray}

For $P_Q$ and $P_\Theta$ at $\Gamma_i$, since $w$ (or $\Phi$) is
antisymmetric,
\begin{eqnarray}
e^{\pi i(\tilde\eta(D_{P_Q})-\tilde\eta(D_{P_\Theta})}={\rm
Pf}(\Phi(P_Q)\Phi^\dag(P_\Theta))=e^{i\sum_\alpha\chi_{k,\alpha}}.
\end{eqnarray}

Because $Q(k=0)$ $=Q(k=\pi)=1$ at $t=\{0,T/2\}$, $P_Q$ is a better
reference projection than $P_C$. The $\eta$-invariants of $P_Q$
vanish, i.e., $\tilde\eta(D_{P_Q})(0)=\tilde\eta(D_{P_Q})(\pi)=0$
at $t=\{0,T/2\}$. This gives rise to, at $t=\{0,T/2\}$,
\begin{eqnarray}
e^{-i\pi[\tilde\eta(D_{P_\Theta(\pi)})
-\tilde\eta(D_{P_\Theta(0)})]}=e^{i\sum_\alpha
(\chi_{\pi,\alpha}-\chi_{0,\alpha})}=\frac{{\rm Pf}(w(\pi))}{{\rm
Pf}(w(0))}
\end{eqnarray}
Taking the logarithm both sides, one has
\begin{eqnarray}
\frac{i}{\pi}\log\frac{{\rm Pf}(w(\pi))}{{\rm
Pf}(w(0))}=(\tilde\eta (D_{P_\Theta(\pi)})
-\tilde\eta(D_{P_\Theta(0)}))~{\rm mod}~2. \label{Pfeta}
\end{eqnarray}
We will use this result later.

\section{Winding number: Kane-Mele's invariant}

In Fu and Kane's work \cite{FK}, Kane-Mele's invariant is defined
by the change of the partial polarization, whose topological
nature is not so manifest and an additional proof was needed. In
this section, we define a topological invariant winding number to
identity Kane-Mele's invariant and show that this winding number
is $\mathcal{Z}_2$-modular to the spectral flow that we defined in Sec.
III. The rigorous mathematical definition of the winding number
can be found in Sec. 6 in Ref. \cite{KL}.

\subsection{Winding Number and Spectral Flow}

In the previous section, we introduce a map $w$ (or $\Phi$) from
an open path $k\in [0,\pi]$ to $U(2N)$. To define a winding number
corresponding to an open path, one needs to introduce an
invertible mapping $f$ so that $f(0)=w(\pi)$ and $f(\pi)=w(0)$.
For a topological insulator whose bulk states are gapped and non-degenerate, one can define an Atiyah-Patodi-Singer
boundary problem so that the edge states are projection away. The
solution of this boundary problem defines an invertible map $f$.

Now we can define a product map $w*f$ which maps
$([0,\pi],[0,\pi])\to (U(2N), U^*(2N))$ in which $U^*(2N)$ are a
subset of $U(2N)$ excluding $2N\times 2N$ matrices which have
eigenvalue $-1$. The product map $w*f$ is a closed path and one
can define a winding number in the conventional way. This winding
number can be thought a winding number of an open path $w$ and is
given by \cite{KL}
\begin{eqnarray}
{\rm wind}(w)&=&\sum_{t=0,T/2}\frac{1}{2\pi i}\biggl[\int_0^\pi \
dk \ {\rm tr}
[w^{-1}(k,t)\frac{d}{dk}w(k,t)]\nonumber\\
&-&{\rm tr}\log w(\pi,t)+{\rm tr}\log w(0,t)\biggr]\nonumber\\
&=&\sum_{t=0,T/2}\frac{1}{2\pi i}\biggl[\int_0^\pi \ dk \ {\rm tr}
[w^{-1}(k,t)\frac{d}{dk}w(k,t)]\nonumber\\
&-&2\log\frac{{\rm Pf}(w(\pi,t))}{{\rm
Pf}(w(0,t))}\biggr]\label{wind}
\end{eqnarray}
After integrating, one arrives at Eq. (\ref{wind1}) up to a
$\mathcal{Z}_2$-modula.

Combining (\ref{wind}), (\ref{Pfeta}) and (\ref{etasf}) together,
one has
\begin{eqnarray}
&&\biggl[SF(D_{P_\Theta(k)})+\frac{1}2 \sum_{t=0,T/2}\int_0^\pi dk
\frac{d\eta(D_{P_\Theta(k)}(t))}{dk}\biggr]~{\rm mod}~2\nonumber\\
&&={\rm wind}(w)+\frac{1}{2\pi}\sum_{t=0,T/2}\int_0^\pi dk ~{\rm
tr}(w^\dag(k,t) i\frac{d}{dk}w(k,t)).\nonumber \\
\end{eqnarray}
Identifying the Berry phases both sides, i.e., the integrations of
continuous functions, we reach the main result 
\begin{eqnarray}
{\rm wind}(w)_{\partial\tau_1}={\rm SF}(D_{P_\Theta}(\tau_1))~{\rm
mod}~2.\label{windsf}
\end{eqnarray}
The spectral flow is integer-valued. This relation shows that the
winding number is indeed  $\mathcal{Z}_2$-valued.

\subsection{Bulk-edge correspondence}

We know explain the physical meaning of Eq. (\ref{windsf}). The
left-hand side is Kane-Mele's invariant which is a topological
invariant. The right hand side says that up to an even number,
Kane-Mele invariant categorizes the spectral flow of the family of
Dirac operators $D_{P(k)}$.

By definition, the spectral flow counts both the discrete zero
modes and the continuous gapless excitations of the system. For
this spin pumping model, the edge of the system is
$S^1\times\{0\}\cup S^1\times\{T/2\}$. In reality, we have an open boundary chain on which the ground state of this spin pumping
model is not degenerate and the bulk states are fully gapped.  Thus, the spectral flow counts the oriented
gapless excitations, i.e., the gapless edge states in the
topological insulator. The
$\mathcal{Z}_2$-modula makes the direction of the edge states
become not important. For the model described in Sec. II, Fu and
Kane have plotted its schematic band structure in Figs. 1 and 2 of
Ref. \cite{FK} in which the spectral flow of $D_{P(k)}(t)$ from
$[0,\pi]$ can be read out and is indeed +1. By considering Kramers partner of the edge states, 
we explains why the parity of the number of the Kramers pairs of edge states exactly
given by Kane-Mele's invariant. In this way, we understand the
edge states distinguish topological insulator from conventional
insulator.

\section{Disorders and interactions}

As well known, weak disorder and weak interaction do not affect
the topological property in quantum Hall effects. Numerical
calculations \cite{shen,xie} and analytic study \cite{groth}
showed that it is also the case for topological insulator. Indeed,
a Kane-Mele-type $\mathcal{Z}_2$-valued topological invariant can
also be defined in a disordered and interacting system
\cite{FK,essin}, in the sprit of the twisted boundary condition
\cite{NT,NWT}. In this section, we show that all results we
obtained in the previous sections may also apply to the disordered
and interacting systems.

 Assume the wave function of the one-dimensional system with disorder obeys the
twisted boundary condition,
\begin{eqnarray}
\psi(x+L)=e^{i\phi}\psi(x).
\end{eqnarray}
This wave function is multi-valued. Making a gauge transformation
\begin{eqnarray}
\chi(\phi,x)=e^{-i\phi x/L}\psi(x),
\end{eqnarray}
the wave function becomes a single-valued one.

The Hamiltonian with disorder then becomes $H\to H(\phi)$  as the
states change. Then under the time reversal transformation,
\begin{eqnarray}
\Theta H(-\phi)\Theta^{-1}=H(\phi)
\end{eqnarray}
Similar to the periodic case, there are four times-reversal points
$\Gamma_i$. The eigen states are $|\chi(\phi,\lambda)\rangle$ with
\begin{eqnarray}
H(\phi)|\chi(\phi,\lambda,{\bf
r})\rangle=\lambda_{\phi}|\chi(\phi,\lambda,{\bf r}) \rangle.
\end{eqnarray}
Making a substitution $(k,t)\to (\phi,t)$ and following a similar
way in the previous sections, we can identify the spectral flow of
the Hamiltonian family $\{H(\phi)\}$ as Kane-Mele-type invariant
after modular an even integer. The correspondence between the edge
states and Kane-Mele-type invariant also holds.

For an interacting system, we can use the twisted boundary
condition to the many body wave function \cite{NT,NWT} and the
above results are still valid.

If there are the strong interaction between the edge modes, it was shown that for a topological
insulator, the edge states may be gapped while the ground state
may become degenerate since the time reversal symmetry is
spontaneously broken \cite{Wu,Xu,FK}. In fact, with a generalized
twisted boundary condition, a topological system has either the
ground state degeneracy or gapless excitations \cite{qwz}.

According to the analysis given in the present work, this result
is quite natural. The spectral flow of the Dirac operator family
is computed by the reduced $\eta$-invariant which counts both the
discrete zero modes and the gapless excitations of the system. The edge interaction, even if  it is very strong, would 
not violate the topological nature of the system and thus the reduced
$\eta$-invariant is not changed even the gapless
excitations are gapped by the spontaneous or perturbative
breaking of the time reversal symmetry because the zero modes, i.e., the ground state degeneracy, compensate the gapless excitations.

On the other hand, the strong bulk interaction may modify the topological classification from $Z_2$ invariant to a $Z_8$ invariant which 
is beyond the free fermion classifications \cite{citi}. To study these strongly interacting systems is not the goal in this work because it may require
more mathematic tools such as the knowledge of the non-linear Dirac equations. 

\section{comments on higher dimensional topological insulators}

The concept of the spectral flow in the present work restricts to
a single-parameter family of the Dirac operators. This confines
the generalization of the study in this work to a higher
dimensional topological insulator. For example, the correspondence
$(k,t)\to (k_x,k_y)$  leads to an understanding to Kane-Mele's
invariant in two spatial dimensions \cite{FK} but the spectral
flow of a two-parameter Dirac operator family needs to be defined.

Mathematically, such a multi-parameter spectral flow  is called
{\it higher spectral flow} which has been introduced in Ref.
\cite{DaiZhang}. Instead of the Atiyah-Patodi-Singer boundary
condition problem, the problem to be solved is a (generalized)
spectral section problem \cite{MP}. It has been shown that there
is an index theorem in which the higher spectral flow may be
expressed by the Chern character in the parameter space
\cite{DaiZhang}, which seems that the equivalence between two
topological invariants proved in Ref. \cite{wqz} is a special
example of this index theorem of the Dirac operator which is
parameterized by the Brillouin zone. To explain these mathematical
results to condensed matter physicists requires many preparations.
We will present them in subsequent works.

\vspace{3mm}

\section{Discussions and conclusions}

We studied the the $\mathcal{Z}_2$-modular relation between
Kane-Mele's invariant in a 1+1-dimensional topological insulator
and the spectral flow of a Dirac operator family of this system.
This relation revealed why and in what case the edge states can
categorize the topological insulators.

In fact, the index theorem associated with an Atiyah-Patodi-Singer
boundary condition problem of the Dirac operator is essential to
characterize the topological nature of a physical system. The
topological insulators can be classified by the topological
index given by the $\eta$-invariant and (higher) spectral flow.
This classification is in fact closely related to other
classifications of Refs. \cite{ram,ktheory}. The random matrix
classification categorizes the topological insulators through the
eigenvalues of the Hamiltonian. In the continuous limit, this
yields to classify the spectrum of the Dirac operator. On the
other hand, the (higher) spectral flow is in fact a representative
element of the K-group in the parameter space \cite{DaiZhang}.
Therefore, our study gives an explicit representation of the
classifications based on K-theory and the random matrix to the
topological insulators.

Recently, there were two papers that relate the $\eta$-invariant in Atiyah-Patodi-Singer index theorems to the topological phases \cite{witten,yonekura}. These works are based on Dai-Freed theorem which simplifies and generalizes the gluing formula
for the $\eta$-invariant \cite{df,BL,buk,woj1,woj2,mull}. The difficulty to calculate the integer contribution in the gluing formula in terms of this theorem was certain non-intrinsic projections are used.  However, the technique used in this work is an intrinsic way to express the integer contribution by the spectral flow of a naturally defined family of self-adjoint operators \cite{KL}.

\noindent{\bf Acknoledgements}  YY is supported in part
by NNSF of China (11474061).

\noindent{\bf Note Added}  The results of this work had been reported in a talk by the second author representing the three authors on the Conference on "Novel Quantum States in Condensed Matter" (Univ. of Chicago Center in Beijing; Sept. 1, 2010), with the title "Atiyah-Patodi-Singer Index Theorem, Spectral Flow and the Topological  Number $Z_2$ for TRI Systems".


\begin{references}
\bibitem{note} Mathematically, these statements are non-trivial
and require many rigorous definitions and proofs. We take them
from mathematical literature, e.g., \cite{KL}.
\bibitem{note2} As we shall see, $\frac{d}{dk}\eta(D_{P(k)})$
corresponds to the Berry connection in Appendix A of Ref.
\cite{FK}.
\end{references}
\end{document}